\documentclass{xray_symp}
\usepackage{graphics,psfig}

\def\etal{{\it et al.}}
\def\msun{$M_{\odot}$}
\def\mdot{$\dot M$}

\def\v7{V751\,Cyg}
\def\rxj0513{RX\,J0513.9--6951}
\def\hd{HD 104994}

\begin{document}

\title{Relation between Supersoft X-ray sources and VY Scl stars}

\author{J. Greiner\inst{1} \and R. Di\,Stefano\inst{2}}  

\institute{Astrophysical Institute Potsdam, An der Sternwarte 16,
             14482 Potsdam, Germany
\and  Harvard-Smithonian Center for Astrophysics, 60 Garden Street, 
               Cambridge, MA 02138, USA}
\maketitle

\begin{abstract}
In a recent ROSAT observation, taken during its optical low state,
 we found that the VY Scl star V751 Cyg 
was a highly luminous source of soft X-rays.
The ROSAT HRI count rate of V751 Cyg was a factor of 10--20 higher than 
upper limits obtained with ROSAT during previous observations, during which
V751 Cyg happened to be in the optical high state. 
Analysis of the X-ray spectrum suggests a very low
temperature ($kT_{\rm bb} \approx$ 15 eV) and a bolometric luminosity
of 5$\times$10$^{36}$ (D/500 pc)$^2$ erg/s. These values are characteristic of
supersoft X-ray sources. Based on the supersoft nature of \v7\ and
the similarity of the variability pattern
of VY Scl stars and the supersoft transient source RX J0513.9--6951
we hypothezise a relation between VY Scl stars and supersoft
X-ray sources. It is conceivable that VY Scl stars are all supersoft binaries
(SSB),
that they form an extension of the previously known class of SSBs 
towards lower white dwarf masses and lower temperatures. In this case
SSBs must be even more numerous than previously thought.
 
\end{abstract}

\section{Introduction}

Supersoft X-ray binaries (SSB) were established as a new class of astronomical
objects during the early years of this decade (Tr\"umper \etal\ 1991,
Greiner \etal\ 1991, Kahabka \& van den Heuvel 1997) and are thought
to contain white dwarfs accreting mass at
rates high enough to allow quasi-steady nuclear burning of the accreted
matter (van den Heuvel \etal\ 1992). The sources are highly luminous 
($L_{bol} \sim 10^{36}-10^{38}$ ergs s$^{-1}$), but since their
 characteristic temperatures are on the order of tens of eV,
much of the energy is radiated in
the far ultraviolet or soft X-ray region of the spectrum, where the
radiation is easily absorbed by the interstellar medium. Because of this, only
2 close-binary Galactic supersoft sources are known 
although there should be about 
1000 in the Milky Way (Di\thinspace Stefano \& Rappaport 1994).
The quest to find new SSBs has inspired several projects.

Here we report on the discovery of a new SSB. We have taken and analyzed
ROSAT data that verify that the binary V751 Cyg is indeed a luminous
supersoft X-ray source. Because of the importance of
being able to identify additional SSBs, we also report on the
reasoning that led us to suspect that V751 Cyg was
likely to be a transient SSB. 

VY Scl stars are a subclass of nova-like, cataclysmic variables which are
bright most of the time, but
 occasionally drop in brightness by several
magnitudes  at irregular intervals (Warner 1995).
The transitions between the brightness levels
occur on  timescales of days to weeks. These variables 
have  large mass transfer
rates \mdot\ (of the order of 10$^{-8}$ \msun/yr),
and thus are thought to be steady accretors with hot disks.
Livio \& Pringle (1994) proposed a model for the group of VY Scl stars in which
the brightness drop is due to a reduced mass transfer rate which in turn
may be caused by a magnetic spot temporarily covering the $L_1$ region.

\v7\  (= EM* LkHA 170 = SVS 1202) was discovered by Martynov (1958),
showing irregular variations and occasional fadings down to about 16$^{\rm m}$
(Martynov \& Kholopov 1958). Now-a-days \v7\ is classified as a 
nova-like variable based on the observed rapid flickering and the complete 
similarity to VY Scl (Robinson \etal\ 1974).
The orbital period of \v7\ is not well known, but estimated to
$P \approx$ 6 hrs (Bell \& Walker 1980).

\section{ROSAT, IUE and optical observations}

\begin{figure*}
  \resizebox{12cm}{!}{\includegraphics{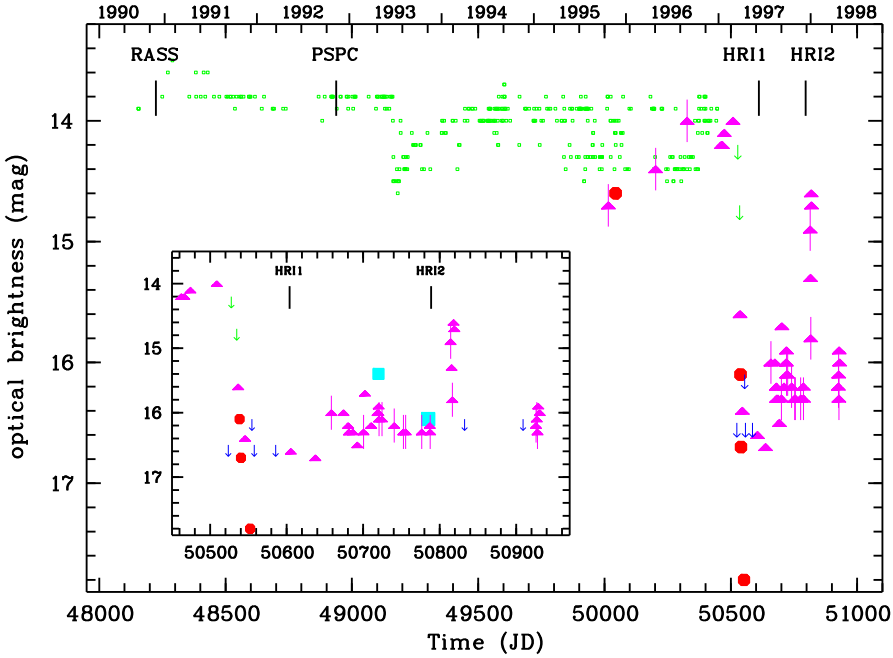}}
  \hfill
  \parbox[b]{55mm}{
\caption[]{Optical light curve of V751 Cyg over the last 8 years:
         small (gray) squares denote measurements as reported to AFOEV 
        and triangles denote measurements as reported to VSNET (uncertain
        measurements are plotted with an error bar while the error of the
        certain measurements is about the size of the symbol sign). 
       Large filled circles
        are CCD measurements of the Ouda team (also taken from VSNET).
        Arrows  denote  upper limits.
         At the top the times of the  ROSAT observations are marked.
         V751 Cyg is detected in X-rays only during the optical
        low state. The inset shows a blow-up of the optical low-state together
       with the times of the two HRI observations. The two squares in the inset
        represent the mean brightness on Sep. 29/30, 1997 as derived from
        our spectra and our Dec. 3 1997 photometry (at the time of the 2nd 
       HRI observation).}
   \label{lc}
   }
\end{figure*}

Full details of the observations of \v7\ will appear elsewhere 
(Greiner \etal\ 1998), 
so we only summarize the relevant information here.
The distinct lightcurve of \rxj0513\ and its similarity
to VY Scl stars led us to decide to monitor 
the light curves of the 14 known VY Scl stars.
When \v7\ started to drop in brightness somewhere between
1 March  and 11 March 1997 (Fig. \ref{lc})
we performed a target-of-opportunity ROSAT HRI observation 
(4660 sec) on 3 June 1997. We discovered a new X-ray source,
RX J2052.2+4419, within 1\arcsec\ of V751 Cyg, at a mean count rate of
0.015 cts/s. During a second ROSAT HRI observation on Dec. 2--8, 1997 the
count rate and X-ray spectrum are nearly identical to the June values.

In contrast, \v7\ was not detected during the ROSAT all-sky survey
on Nov. 19/20, 1990 giving a 3$\sigma$ upper limit of 0.019 cts/s in the PSPC.
In addition, it was also not detected during a serendipituous pointing
on Nov. 11, 1992 providing an upper limit of 0.0058 cts/s in the PSPC.
On both occasions \v7\ was in its optical bright state. We therefore find 
evidence for an  anti-correlation of optical and X-ray intensity in \v7.

\begin{figure}
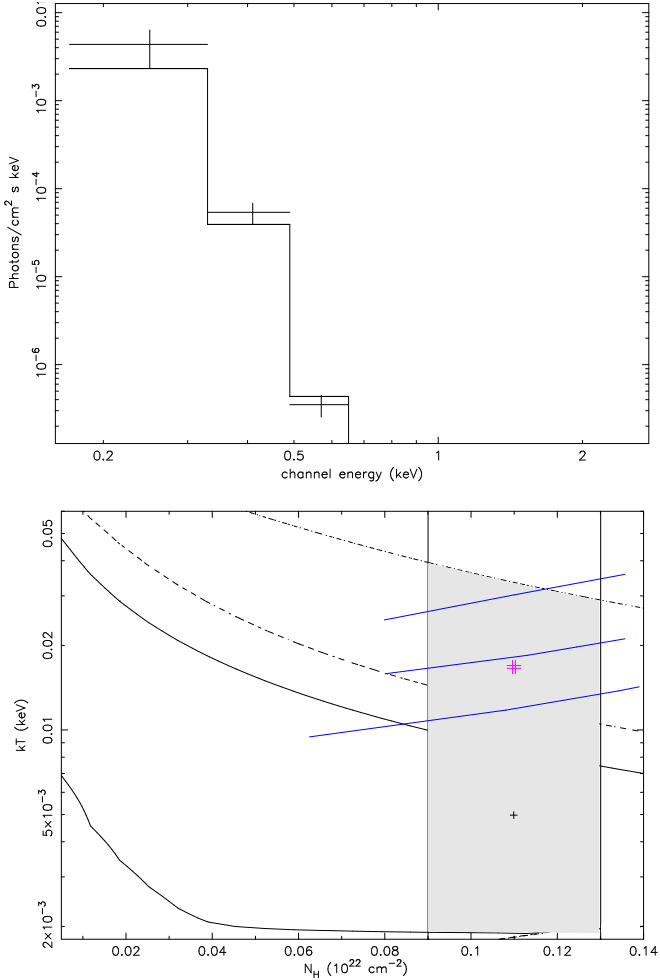

    \vbox{\psfig{figure=v751cyg_ufs.ps,width=8.8cm,%
        bbllx=2.4cm,bblly=2.4cm,bburx=18.cm,bbury=13.8cm,clip=}}\par
     \vspace*{.3cm}
    \vbox{\psfig{figure=v751cyg_cont.ps,width=8.8cm,%
        bbllx=0.9cm,bblly=1.4cm,bburx=16.8cm,bbury=12.8cm,clip=}}\par
       \vspace*{-0.2cm}
    \caption[pha]{Spectral fit result of the June 1997 observation
         of V751 Cyg:
        {\bf Top:} The photon spectrum deconvolved with a blackbody model with
         $N_{\rm H}$ fixed at the value  derived from the IUE spectrum.
         {\bf Bottom:}  Confidence contours of the blackbody fit in the
         $kT-N_{\rm H}$ plane: 68\% (solid line),
         90\% (long-dashed line), 99\% (short-dashed line).
         The two vertical lines denote the $N_{\rm H}$ range allowed
         by the IUE data ((1.1$\pm$0.2)$\times$10$^{21}$ cm$^{-2}$),
        and the vertical extent of the hatched region marks the 99\% confidence
         region of the blackbody temperature.
         The three solid  lines crossing the hatched region
        mark contours of constant luminosity (at an assumed distance of 500 pc)
         of 10$^{34}$, 10$^{36}$ and 10$^{38}$ erg/s (from top to bottom),
         respectively.
   The cross denotes the best fit-value of $kT$=5\,eV and the double cross the
   parameter pair used in the  discussion (see sect. 2).
         }
      \label{ufs}
      \vspace{-0.2cm}
   \end{figure}

Using a new method (Prestwich \etal\ 1998) to extract reliable spectral  
information from HRI data, we crafted a response matrix which takes
into account both the
gain state of the detector at the time of the observation and the ``wobble''
of the HRI during the detection.       
Fits using this response matrix to all the source photons 
of V751 Cyg show that simple black-body 
models with kT of a few tens of eV 
are consistent with the data, 
whereas higher temperature models (0.5 keV) can be ruled out (Fig. \ref{ufs}).

An IUE observation performed in 1985 (also during an optical high state)
was used to derive the extinction towards \v7\ based on
the broad absorption centered at  2200 \AA.
The  best result is obtained with a value of $E(B-V)=0.25\pm0.05$.
With the intrinsic color being near zero, this implies a visual
extinction of $A_{\rm V} = 0.82\pm0.17$. Using a mean extinction law
of $A_{\rm V}$ = 1.9 mag/kpc (Allen 1973) we derive
a distance to \v7\ of 430$\pm$100 pc, while using that
from  Neckel \& Klare (1980) gives $d$=610$\pm$30 pc.
Using the absorbing column derived from IUE,
$N_{\rm H}$ = 1.1$\times$10$^{21}$ cm$^{-2}$, the best-fit blackbody model
gives $kT = 5$ eV and extremely high luminosity, so we adopt
$kT = 15^{+15}_{-10}$ eV in the following (see Fig. \ref{ufs}).

At this temperature, the bolometric luminosity on 3 June 1997 is 
6.5$\times$10$^{36}$ (D/500 pc)$^2$ erg/s. Thus, during its optical low state,
\v7\ was emitting soft X-rays with a temperature and luminosity
which confirm that it is a transient supersoft X-ray source.

\section{V751 Cyg as a supersoft X-ray binary}

The following picture emerges.
During optical low states \v7\ exhibits transient soft X-ray emission
thus revealing itself as a  supersoft X-ray binary.
The appearance of \ion{He}{ii} 4686 \AA\ emission in optical spectra taken
in Sep. 1997 also indicates the presence 
of $>$54 eV photons.
V751 Cyg, like the other members of the VY Scl star group,
accretes at a few times 10$^{-8} M_\odot$ yr$^{-1}$. If the mass of the
white dwarf in \v7\ is small, this may allow nuclear burning, as the
high X-ray luminosity suggests. It is worth emphasizing that recent
calculations of hydrogen-accreting carbon-oxygen
white dwarfs have shown that the accretion rate for low mass white dwarfs
(0.5--0.6 \msun) can be as low as 1--3$\times$10$^{-8}$ \msun/yr
(Sion \& Starrfield 1994, Cassisi \etal\ 1998) while still maintaining
shell burning (consistent with Fujimoto 1982).
The \v7\ values of $M_{\rm V}^{\rm max} = 3.9$  and
${\log \Sigma} = (L_x/L_{\rm Edd})^{1/2} P_{\rm orb}^{2/3} (hr) = -0.23$ are
consistent, within the uncertainties of $L_x$ and $P_{\rm orb},$ with the
relation $M_{\rm V} = 0.83(\pm 0.25) - 3.46 (\pm 0.56) \log \Sigma$
found for 5 SSB (van Teeseling \etal\ 1997) implying
that, if nuclear burning is the correct interpretation
of the X-ray flux during the optical low state,
then nuclear burning may continue during the optical high state.

There is ample evidence in some VY Scl stars that during the
optical low state the accretion disk has vanished. Though we have no 
optical observations to provide direct
evidence for this in \v7, the disk is certainly optically thin, thus
drastically reducing the efficacy of reprocessing. This would cause
some difference between the optical spectra of \v7\ and the SSBs 
in which the optical light is dominated by reprocessed light from the 
accretion disk (Popham \& DiStefano 1996).

Our discovery that \v7\ is a transient supersoft X-ray source arose from the
similarity in the optical light curve of RX J0513.9--6951 and VY Scl stars.
\rxj0513 (Schaeidt \etal\ 1993, Pakull \etal\ 1993) shows $\sim$4 week 
optical low states which are accompanied by luminous supersoft X-ray 
emission (Reinsch \etal\ 1996, Southwell \etal\ 1996). It is generally 
assumed that the
white dwarf accretes at a rate slightly higher than the burning rate,
and thus is in an inflated state during the optical high state.
Changes in the irradiation of the disk caused by the expanding/contracting
envelope around the white dwarf have been proposed as explanation for the
1 mag intensity variation in \rxj0513\ (Reinsch \etal\ 1996, 
Southwell \etal\ 1996). 
Note, that the white dwarf itself varies drastically as it
expands/contracts, and in fact a flaring disk had to be assumed for
\rxj0513\ to reduce the theoretically possible amplitude down to only 1 mag
(Hachisu \& Kato 1998).

The explanation of the X-ray/optical variability of \v7\ could be
similar to RX J0513.9--6951
(Pakull \etal\ 1993, Reinsch \etal\ 1996, Southwell \etal\ 1996):
\mdot variations change both the photospheric radius and
the disk spectrum.
If the white dwarf has a small mass, than photospheric radius expansion
is reached  at 1$\times$10$^{-7}$ \msun/yr (Cassisi \etal\ 1998).

A major difference between the optical light curves of \rxj0513\ and
VY Scl stars is  the amplitude between low and high states, i.e. 1 mag
(\rxj0513) versus 3--6 mag for VY Scl stars (4 mag for \v7).
Note that the $\approx$15 eV blackbody model derived as the best fit for \v7\
corresponds to a m$_{\rm V} \approx$20 mag, i.e. several magnitudes fainter
than the observed optical low-state intensity (Fig. \ref{lc}).
Indeed, an amplitude of 4 mag can be easily accommodated by a white dwarf when
expanding from R$_{\rm WD}$ to 5 R$_{\rm WD}$. Thus, the observed large
amplitudes in VY Scl stars
could be due to a combination of both the disk disappearance and the white
dwarf contraction.

\section{Are other nova-like variables also SSBs?}

\subsection{VY Scl stars}

The discovery of  luminous, supersoft X-ray emission during the optical low
state of \v7\ naturally leads to the question of whether other
VY Scl stars may also be SSBs.
It turns out that the basic properties of VY Scl stars correspond
surprisingly well to an extension of the SSB class (see Tab. \ref{com}): 
(i) In all thouroughly observed VY Scl stars, optical emission line studies 
indicate that the {\it donor} has a mass smaller than $0.5$ \msun.
In addition, the {\it mass ratio} (between the donor and the
white-dwarf accretor) seems to be close to unity in some systems.
(ii) The {\it white dwarf mass} in VY Scl systems may be systematically smaller
than generally considered in models for SSBs.
We therefore may be seeing an extension 
of the CO-nuclear-burning white dwarf scenario which has formed the basic model for most SSBs so far.
The low white dwarf mass may
lead to a  larger effective radii (Vennes \etal\ 1995) and thus lower
temperatures.
(iii) The orbital periods of VY Scl stars range from $\sim 3.2$ hours to $6$
hours, with the longest period associated with V751 Cyg. These periods
are compatible with
the orbital periods of some other SSBs, most notably 1E 0035.4--7230
with its 4.1 hr orbital period (Schmidtke \etal\ 1996). However, such 3--4 hr
orbital periods are significantly lower than those required
in the canonical van den Heuvel \etal\ (1992) model for supersoft sources
in which a possibly slightly evolved donor more massive than the white 
dwarf provides a mass transfer
on a thermal timescale. However, it has recently been shown that the strong
X-ray flux in supersoft sources should excite a strong wind
($\dot M_{\rm wind} \sim 10^{-7}$ \msun/yr) from the irradiated companion
which in short-period binaries would be able to drive Roche lobe
overflow at a rate comparable to $\dot M_{\rm wind}$
(van Teeseling \& King 1998).
(iv) The value of $\dot m$ typically adopted in VY Scl systems (Warner 1987) is
$\dot m \sim 10^{-8} M_\odot$ yr$^{-1}$.
Interestingly enough, these values
may be compatible with quasi-steady burning on white dwarfs with
the low masses that seem to be indicated in some VY Scl systems. 

Thus, the conjecture that all VY Scl stars are SSBs is viable. Should it be 
verified, even for a subset of systems, then VY Scl stars may represent an
extension of the class of SSBs  as discussed above and summarized 
in Tab. \ref{com}. 
Even if the underlying mechanisms in SSB and \v7\ 
are completely different, the somewhat similar anticorrelations between 
X-ray and optical have proved useful to identify \v7\ as a transient
supersoft X-ray source.

\subsection{V Sge stars}

It has recently been suggested (Steiner \& Diaz 1998; Patterson \etal\ 1998)
V Sge (and possibly also WX Cen, V617 Sge and \hd) have
properties very similar to SSBs.
These suggestions are based on the following characteristics,
shared by these four stars but rare or even absent among canonical
cataclysmic variables:
(1) the presence of both \ion{O}{VI} and \ion{N}{V} emission lines,
(2) a \ion{He}{II} $\lambda 4686$/H$\beta$ emission line ratio $\ga 2$,
(3) rather high absolute magnitudes and very blue colours, and
(4) orbital lightcurves which are characterized by a wide and deep eclipse.
A detailed comparison of the optical states of V Sge and archival ROSAT
observations has shown that during optical bright states, V~Sge is a faint 
hard X-ray source, while during optical faint states ($V \ga 12$ mag), 
V~Sge is a `supersoft' X-ray source (Greiner \& Teeseling 1998). 
Spectral fitting confirms that V~Sge's X-ray properties during
its soft X-ray state may be similar to those of supersoft X-ray
binaries, although a much lower luminosity cannot be excluded.
It is possible to explain
the different optical/X-ray states of V~Sge by a variable amount of
extended uneclipsed matter,
which during the optical bright states contributes significantly
to the optical flux and completely absorbes the soft X-ray component
(Greiner \& Teeseling 1998).
We note that our strategy of finding new, transient SSB by measuring the 
X-ray emission during optical low states would have worked also for V Sge.

\begin{table}[th]
    \caption{Comparison of SSB and VY Scl group properties}
    \vspace{-0.2cm}
    \begin{tabular}{ccc}
    \hline 
    \noalign{\smallskip}
     & SSBs & VY Scl stars \\
    \noalign{\smallskip} 
    \hline 
    \noalign{\smallskip}
     Mass of WD ($M_{\odot}$)    & $\sim$ 1    & $\sim$0.5 \\
     Mass of Donor (\msun) & $\sim$ 1 -- 2 & $\sim$0.5 -- 0.7 \\
     Period (hrs)          & 6 -- 70       & 3 -- 6 \\
     kT (eV)               & 20 -- 50      & 10 -- 20 \\
     Accretion rate ($M_{\odot}$/yr) & 10$^{-7}$  & 10$^{-8}$ \\
     M$_{\rm V}$ (mag)     &  --2 -- +1    &  3 -- 5 \\
     log L (erg/s)         & 37--38      &  36 \\
     Number in Galaxy (obs) & 2          & 15 \\     
     Number in Galaxy (mod) & ~~~1000--3000~~~ & ?? \\
   \hline
   \end{tabular}
   \label{com}
\end{table}

{\em \noindent Although there is certainly a lot of work to be done we are 
encouraged that the method of finding new SSB based on variability
analysis seems capable to expand the realm of known SSB.}

\begin{acknowledgements}
JG thanks Prof. J. Tr\"umper for granting ROSAT TOO time which made this
investigation only possible. We are extremely grateful to J. Mattei for 
providing very useful AAVSO data, A. Prestwich for calculating the
HRI detector response matrix, R. Gonz\'alez-Riestra for analysing the
IUE spectrum and G. Tovmassian for obtaining optical observations.
JG is supported by the German 
BMBF/DLR under contract No. FKZ 50 QQ 9602 3.
The ROSAT project is supported by BMBF/DLR and the Max-Planck-Society.
\end{acknowledgements}

\end{document}